\documentstyle[12pt]{article}
\date{}
\author{Valerii Dryuma\thanks{Work supported in part by MURST, Italy}\\[5mm]
{\it Institute of Mathematics and Informatics, AS RM,}\\[3mm]
{\it 5 Academiei Street, 2028 Kishinev, Moldavia},\\[3mm]{\it e-mail: 
valery@gala.moldova.su} }
\title{ON THE RIEMANNIAN AND EINSTEIN-WEYL
GEOMETRY IN THEORY
OF THE SECOND ORDER ORDINARY DIFFERENTIAL EQUATIONS}

\newtheorem{pr}{Proposition}
\newtheorem{rem}{Remark}
\begin{document}
\maketitle
\date{}
\maketitle
\begin{abstract}
Some properties of the 4-dim Riemannian spaces with metrics
$$
ds^2=2(za_3-ta_4)dx^2+4(za_2-ta_3)dxdy+2(za_1-ta_2)dy^2+2dxdz+2dydt
$$
 associated with the second order nonlinear differential equations
$$
y''+a_{1}(x,y){y'}^3+3a_{2}(x,y){y'}^2+3a_{3}(x,y)y'+a_{4}(x,y)=0
$$
with arbitrary coefficients $a_{i}(x,y)$ and 3-dim Einstein-Weyl
spaces connected with dual equations
$$
b''=g(a,b,b')
$$
where the function $g(a,b,b')$ satisfied the partial differential equation
$$
g_{aacc}+2cg_{abcc}+2gg_{accc}+c^2g_{bbcc}+2cgg_{bccc}+
g^2g_{cccc}+(g_a+cg_b)g_{ccc}-4g_{abc}-
$$
$$
 -4cg_{bbc} -cg_{c}g_{bcc}-
3gg_{bcc}-g_cg_{acc}+ 4g_cg_{bc}-3g_bg_{cc}+6g_{bb} =0
$$
are considered.
\end{abstract}

\section
{Introduction}

The  second order ODE's of the type
$$
y''+a_1(x,y){y'}^3+3a_2(x,y){y'}^2+3a_3(x,y)y'+
a_4(x,y)=0  \eqno (1)
$$
are connected with  nonlinear dynamical systems in the form
$$
\frac{dx}{dt}= P(x,y,z,\alpha_i),\quad \frac{dy}{dt}= Q(x,y,z,\alpha_i),
\quad \frac{dz}{dt}= R(x,y,z,\alpha_i),
$$
where $\alpha_i$ are parameters.

For example the Lorenz system
$$
\dot X=\sigma(Y-X),\quad \dot Y=rX-Y-ZX,\quad \dot Z=XY-bZ
$$
having  chaotic properties at some values of parameters
is equivalent to the equation
$$
y''-\frac{3}{y}{y'}^2+(\alpha y - \frac{1}{x})y'+\epsilon xy^4-\beta x^3 y^4-
\beta x^2y^3-\gamma y^3 + \delta \frac{y^2}{x}=0, \eqno (2)
$$
where
$$
\alpha=\frac{b+\sigma+1}{\sigma},\quad \beta=\frac{1}{\sigma^2},
\quad \gamma=\frac{b(\sigma+1)}{\sigma^2},\quad \delta=\frac{(\sigma+1)}{\sigma},
\quad \epsilon=\frac{b(r-1)}{\sigma^2},
$$
and for investigation of its properties  the theory of
invariantes was first used in [1--5].

According to this theory [6--10] all equations of
type (1) can be devided in  two different classes

I. \ $\nu_5=0$,

II. \ $\nu_5\neq0$.

   Here the value $\nu_5$ is the expression of the form
$$
\nu_5=L_2(L_1L_{2x}-L_2L_{1x})+L_1(L_2L_{1y}-L_1L_{2y})-a_1L_1^3+3a_2L_1^
2L_2-3a_3L_1L_2^2+a_4L_2^3\,
$$
then $ L_1,L_2 $ are defined by fomulas
$$
L_1=\frac{\partial}{\partial y}(a_{4y}+3a_2a_4)-\frac{\partial}{\partial x}
(2a_{3y}-a_{2x}+a_1 a_4)-3a_3(2a_{3y}-a_{2x})-a_4a_{1x} ,
$$
$$
L_2=\frac{\partial}{\partial x}(a_{1x}-3a_1a_3)+\frac{\partial}{\partial x}
(a_{3y}-2a_{2x}+a_1a_4)-3a_2(a_{3y}-2a_{2x})+a_1a_{4y} .
$$

For the equations with condition $\nu_5 =0$
R. Liouville  discovered  the series of semi-invariants
starting from :
$$
w_1={1\over L_1^4}\left[L_1^3(\alpha' L_1-\alpha'' L_2)+R_1(
L_1^2)_x-L_1^2R_{1x}+L_1R_1(a_3L_1-a_4L_2)\right] ,
$$
where
$$
 R_1= L_1L_{2x}-L_2L_{1x}+a_2 L_1^2-2 a_3 L_1L_2 +a_4 L_2^2
$$
or
$$
w_2={1\over L_2^4}\left[L_2^3(\alpha' L_2-\alpha L_1)-R_2(
L_2^2)_y+L_2^2R_{2y}-L_2R_2(a_1L_1-a_2L_2)\right] ,
$$
where
$$
 R_2= L_1L_{2y}-L_2L_{1y}+a_1 L_1^2-2 a_2 L_1L_2 +a_3 L_2^2
$$
and
$$
 \alpha=a_{2y}-a_{1x}+2(a_1 a_3-a_2^2), \quad
 \alpha'=a_{3y}-a_{2x}+a_1 a_4-a_2a_3,
$$
$$
 \alpha''= a_{4y}-a_{3x}+2(a_2 a_4-a_3^2).
$$
It has the form
$$
w_{m+2}=L_1\frac{\partial w_m}{\partial y} -L_2\frac{\partial w_m}
{\partial x}+mw_m(\frac{\partial L_2}{\partial x}-\frac{\partial L_1}
{\partial y}).
$$
  In case $w_1=0$ there are another series of semi-invariants 
$$    
i_{2m+2}=L_1\frac{\partial i_{2m}}{\partial y} -L_2\frac{\partial i_{2m}}
{\partial x}+2mi_{2m}(\frac{\partial L_2}{\partial x}-\frac{\partial L_1}
{\partial y}).
$$
where
$$  
i_2={3R_1\over L_1}+{\partial L_2\over\partial x}-{\partial L_1\over  
\partial y} .  
\label{i2} 
$$  
and corresponding sequence for absolute invariants
$$
j_{2m}=\frac{i_{2m}}{i_2^{m}}.
$$     
 
    In case $\nu_5\neq0$  the semi-invariants have the form
$$
\nu_{m+5}=L_1\frac{\partial \nu_{m}}{\partial y} -L_2\frac{\partial \nu_{m}}
{\partial x}+m\nu_{m}(\frac{\partial L_2}{\partial x}-\frac{\partial L_1}
{\partial y}).
$$
and corresponding serie of absolute invariants
$$
[5t_{m}-(m-2)t_7t_{m-2}]\nu_5^{2/5}=5(L_1\frac{\partial t_{m-2}}{\partial y} -
L_2\frac{\partial t_{m-2}}{\partial x})
$$
where
$$
t_m=\nu_m \nu_5^{-m/5}
$$

\section{Riemannian spaces in theory of ODE's}

Here we present the construction of the Riemannian spaces
connected with the equations (1).

    We start from the equations of geodesical lines
of two-dimensional space $A_2$ equipped with affine (or Riemannian) connection.
They have the form
$$
\ddot x +\Gamma^1_{11} \dot x^2 +2 \Gamma^1_{12} \dot x \dot y+
\Gamma^1_{22} \dot y^2 =0,
$$
$$
\ddot y +\Gamma^2_{11} \dot x^2 +2 \Gamma^2_{12} \dot x \dot y +
\Gamma^2_{22} \dot y^2 =0.
$$
 This system of equations is equivalent to one equation
$$
  y''-\Gamma^1_{22}{y'}^3+(\Gamma^2_{22}-2\Gamma^1_{12}){y'}^2+
(2\Gamma^2_{12}-\Gamma^1_{11})y'+\Gamma^2_{11}=0
$$
of type (1) but with special choice of coefficients  $a_i(x,y)$.

The equations (1) with arbitrary coefficients $a_i(x,y)$ may be considered
as equations of geodesics of 2-dimensional space $A_2$
$$
\ddot x -a_3 \dot x^2 -2a_2 \dot x \dot y-a_1 \dot y^2 =0,
$$
$$
\ddot y +a_4 \dot x^2 + 2a_3 \dot x \dot y + a_2 \dot y^2 =0
$$
equipped with the projective connection  with components
$$
\Pi_1=\left |\begin{array}{cc}
-a_3 & -a_2 \\ 
a_4 & a_3
\end{array} \right |,
\quad
\Pi_2=\left |\begin{array}{cc}
-a_2 & -a_1 \\ 
a_3 & a_2
\end{array} \right |.
$$

The curvature tensor of this type of connection  is
$$
R_{12}= \frac{\partial \Pi_2}{\partial x}-
\frac{\partial \Pi_1}{\partial y}+\left[\Pi_1,\Pi_2\right]
$$
and it has the components
$$
R^{1}_{112}=a_{3y}-a_{2x}+a_1a_4-a_2a_3=\alpha',\quad
R^{1}_{212}=a_{2y}-a_{1x}+2(a_1a_3-a_{2}^2)=\alpha,
$$
$$
R^{2}_{112}=a_{3x}-a_{4y}+2(a_{3}^2-a_2a_4)=-\alpha'',\quad
R^{2}_{212}=a_{2x}-a_{3y}+a_3a_2-a_1a_4=-\alpha'.
$$

For construction of the Riemannian space connected with the equation
of type (1) we use the notice of Riemannian extension  $W^4$ of space $A_2$
with connection $\Pi^k_{ij}$  [12] . The corresponding  metric is
$$
ds^2=-2\Pi^k_{ij}\xi_k dx^i dx^j+2d\xi_idx^i
$$
and in our case it takes the following form ($\xi_1=z, \xi_2=\tau$)
$$
ds^2=2(z a_3-\tau a_4)dx^2+4(z a_2-\tau a_3)dx dy +2(z a_1-\tau a_2)dy^2
+2dx dz +2dy d \tau. \eqno(3)
$$
So, it is possible to formulate the following statement

\begin{pr}

    For a given equation of type (1) there is exists the Riemannian space with 
metric (3) having integral curves of such type of equation as part of its
geodesics.
\end{pr}

Really, the  calculation of geodesics of the space $W^4$ with the metric (3)
lead to the system of equations
$$
\frac{d^2 x}{ds^2} -a_3 \left(\frac{d x}{ds}\right)^2-
2 a_2\frac{d x}{ds}\frac{d y}{ds}-a_1\left(\frac{d y}{ds}\right)^2=0,
$$
$$
\frac{d^2 y}{ds^2} +a_4 \left(\frac{d x}{ds}\right)^2+
2 a_3\frac{d x}{ds}\frac{d y}{ds}+a_2\left(\frac{d y}{ds}\right)^2=0,
$$
$$
\frac{d^2 z}{ds^2} +[z(a_{4y}-\alpha'')-
\tau a_{4x}] \left(\frac{dx}{ds}\right)^2+
2[za_{3y}- \tau (a_{3x}+\alpha'')] \frac{dx}{ds}\frac{dy}{ds}+
$$
$$
+[z(a_{2y}+\alpha)- \tau (a_{2x}+2\alpha')]\left(\frac{d y}{ds}\right)^2+
2a_3 \frac{dx}{ds}\frac{dz}{ds}-2a_4 \frac{dx}{ds}\frac{d \tau}{ds}+
2a_2 \frac{dy}{ds}\frac{dz}{ds}-2a_3 \frac{dy}{ds}\frac{d \tau}{ds}=0,
$$
$$
\frac{d^2 \tau}{ds^2}+[z(a_{3y}-2\alpha')- \tau (a_{3x}-\alpha'')]
\left(\frac{dx}{ds}\right)^2+
2[z(a_{2y}-\alpha)-\tau a_{2x}] \frac{dx}{ds}\frac{dy}{ds}+
$$
$$
+[za_{1y}-\tau(a_{1x}+\alpha)] \left(\frac{dy}{ds}\right)^2+
2a_2 \frac{dx}{ds}\frac{dz}{ds}-2a_3 \frac{dx}{ds}\frac{d \tau}{ds}+
2a_1 \frac{dy}{ds}\frac{dz}{ds}-2a_2 \frac{dy}{ds}\frac{d \tau}{ds}=0.
$$
This system of equations has the integral
$$
2(za_3-\tau a_4){\dot x}^2+4(za_2-\tau a_3)\dot x \dot y+
2(za_1-\tau a_2){\dot y}^2+2 \dot x \dot z +2 \dot y \dot \tau=1.
$$
Remark that first two equations of this system are equivalent to the
equation (1).

  Thus, we construct the four-dimensional Riemannian space with metric (3) and
connection
$$
\Gamma_1=\left | \begin{array}{cccc}
-a_3 & -a_2 & 0 & 0 \\
a_4 & a_3 & 0 & 0 \\
z(a_{4y}- \alpha'')-\tau a_{4x} & z a_{3y}- \tau(a_{3x}+\alpha'') & a_3 & -a_4 \\
z(a_{3y}- 2 \alpha')-\tau (a_{3x}-\alpha'') & z(a_{2y}-\alpha)- \tau a_{2x} & a_2 & -a_3 \cr
\end{array} \right |,
$$
$$
\Gamma_2=\left |\begin{array}{cccc}
-a_2 & -a_1 & 0 & 0 \\
a_3 & a_2 & 0 & 0 \\
z a_{3y}- \tau(a_{3x}+\alpha'') & z(a_{2y}+\alpha)- \tau(a_{2x}+2\alpha') & a_2 & -a_3 \\
z(a_{2y}- \alpha)-\tau a_{2x} & z a_{1y}- \tau(a_{1x}+\alpha) & a_1 & -a_2
\end{array} \right |, \quad
$$
$$
\Gamma_3=\left |\begin{array}{cccc}
0 & 0 & 0 & 0 \\
0 & 0 & 0 & 0 \\
a_{3} & a_2 & 0 & 0 \\
a_{2} & a_{1} & 0 & 0
\end{array} \right |, \quad
\Gamma_4=\left | \begin{array} {cccc}
0 & 0 & 0 & 0 \\
0 & 0 & 0 & 0 \\
-a_{4} & -a_3 & 0 & 0 \\
-a_{3} & -a_{2} & 0 & 0
\end{array} \right |.
$$

The curvature tensor of this metric has the form
$$
R^1_{112}=-R^3_{312}=-R^2_{212}=R^4_{412}= \alpha', \quad
R^1_{212}=-R^4_{312}= \alpha, \quad
R^2_{112}=-R^3_{412}=-\alpha'',
$$
$$
R^1_{312}= R^1_{412}= R^2_{312}=R^2_{412}=0,
$$
$$
R^3_{112}=2z(a_{2}\alpha''-a_{3}\alpha')+2\tau(a_4 \alpha'-a_3 \alpha''),
$$
$$
R^4_{212}=2z(a_{3}\alpha'-a_{2}\alpha)+2\tau(a_3 \alpha-a_2 \alpha'),
$$
$$
R^3_{212}=z(\alpha_x-\alpha'_y+ a_1\alpha''-a_3 \alpha)+\tau(\alpha''_y-
\alpha'_x+a_4\alpha-a_2\alpha''),
$$
$$
R^4_{112}=z(\alpha'_y-\alpha_x+a_1 \alpha''-a_3 \alpha)+\tau(\alpha'_x-
\alpha''_y+a_4\alpha-a_2\alpha''),
$$

Using the expessions for components of projective curvature of
space $A_2$
$$
L_1= \alpha''_y- \alpha'_x+a_2 \alpha''+a_4 \alpha-2a_3 \alpha',
$$
$$
L_2= \alpha'_y- \alpha_x+a_1 \alpha''+a_3 \alpha-2a_2 \alpha',
$$
they  can be presented in form
$$
R^4_{112}=z(L_2+2a_2 \alpha'-2a_3 \alpha)-\tau(L_1+2a_3 \alpha' -
2a_4\alpha),
$$
$$
R^3_{212}=z(-L_2+2a_1 \alpha''-2a_2 \alpha')+\tau(L_1+2a_3 \alpha' -
-2a_2 \alpha''),
$$
$$
R_{13}=\left |\begin{array}{cccc}
0 & 0 & 0 & 0 \\
0 & 0 & 0 & 0 \\
0 & -\alpha' & 0 & 0 \\
\alpha' & 0 & 0 & 0
\end{array} \right |,\quad
R_{14}=\left |\begin{array}{cccc}
0 & 0 & 0 & 0 \\
0 & 0 & 0 & 0 \\
0 & \alpha'' & 0 & 0 \\
-\alpha'' & 0 & 0 & 0
\end{array} \right |, \quad
$$
$$
R_{23}=\left |\begin{array}{cccc}
0 & 0 & 0 & 0 \\
0 & 0 & 0 & 0 \\
0 & -\alpha & 0 & 0 \\
\alpha & 0 & 0 & 0
\end{array} \right | ,\quad
R_{24}=\left |\begin{array}{cccc}
0 & 0 & 0 & 0 \\
0 & 0 & 0 & 0 \\
0 & \alpha' & 0 & 0 \\
-\alpha' & 0 & 0 & 0
\end{array} \right |,
$$
$$
R^i_{j34}=0.
$$ 

     The Ricci tensor $R_{ik}=R^l_{ilk}$ of the space $W^4$ has the components
$$
R_{11}=2 \alpha'',\quad R_{12}=2 \alpha',\quad R_{22}=2 \alpha,
$$
and scalar curvature $R=g^{in}g^{km}R_{nm}$ of the space $W^4$ is
$R=0$. Now we introduce the tensor
$$
L_{ijk}=\nabla_k R_{ij}-\nabla_j R_{ik}=R_{ij;k}-R_{ik;j}.
$$
It has the following components
$$
L_{112}=-L_{121}=2L_1, \quad L_{221}=L_{212}=-2L_2
$$
and with help of them the invariants of equations (1) may be constructed
using the covariant derivations of the curvature tensor and the values
$L_1, L_2$.

     The Weyl tensor of the space $M^4$ is
$$
C_{lijk}= R_{lijk}+\frac{1}{2}(g_{jl} R_{ik}+g_{ik} R_{jl}-
g_{jk} R_{il} - g_{il} R_{jk})+\frac{R}{6}(g_{jk} g_{il}- g_{jl} g_{ik}).
$$
It has only one component
$$
C_{1212}=tL_1-zL_2.
$$

   Using the components of  Riemannian tensor which are not zero
$$
R_{1412}=\alpha'',\quad R_{2412}=\alpha',\quad R_{2312}=-\alpha, R_{3112}=\alpha',
$$
$$
  R_{1212}=z(\alpha_x-\alpha'_y+a_1\alpha''-2a_2\alpha'+a_3\alpha)+
t(\alpha''_y-\alpha'_x-a_4\alpha+2a_3\alpha'-a_3\alpha'')
$$
the equation 
$$
\left |\begin{array}{c}
 R_{ab}-\lambda g_{ab}
\end{array} \right |=0
$$
may be investigate.

      With help of the Weyl tensor the properties of invariantes of the space $M^4$ 
may be investigated. In particular, all invariants of the second order
of the space $M^4$ are equal to zero.

\begin{rem}
   
    The spaces with metrics (3) are flat for the equations (1) with conditions
$$
\alpha=0, \quad \alpha'=0, \quad \alpha''=0,
$$ 
on coefficients $a_i(x,y)$. 

Such type of equations have the components of projective curvature  
$$
L_1=0, \quad L_2=0 
$$
and they are reduced to the the form $y''=0$ with help of points 
transformations.

   On the other hand there are examples of equations (1) with conditions
$L_1=0, \quad L_2=0$ but 
$$
\alpha \neq 0, \quad \alpha'\neq 0, \quad \alpha''\neq0.
$$ 
For such type of equations the curvature of corresponding Riemannian spaces
is not equal to zero.
  
   In fact, the equation
$$
y''+2e^{\varphi}y'^3-\varphi_y y'^2+\varphi_x y'-2e^{\varphi}=0
$$
where the function $\varphi(x,y)$ is solution of 
the  Wilczynski-Tzitzeika nonlinear equation integrable by the Inverse
Transform Mehod. 
$$
\varphi_{xy}=4e^{2 \varphi}-e^{-\varphi}.
$$
has conditions  $L_1=0, \quad L_2=0$ but 
$$
\alpha \neq 0, \quad \alpha'\neq 0, \quad \alpha''\neq0.
$$
\end{rem}  
\begin{rem}
    
     The properties of the Riemann spaces with metrics (3) for the equations
 (2) with chaotical behaviour at the values of coefficients  
$(\sigma=10,\quad b=8/3,\quad r > 24)$ have a special interest. 
The Riemannian spaces of this type are characterized by the special 
conditions on the curvature tensor and its invariantes. 

    The study of the geodesic deviation equation
$$
\frac{d^2 \eta^i}{ds^2}+2\Gamma^{i}_{lm}\frac{dx^m}{ds}\frac{d\eta^l}{ds}+
\frac{\partial \Gamma^{i}_{kl}}{\partial x^j}\frac{dx^k}{ds}
\frac{dx^l}{ds}\eta^j=0
$$
also may be usefull for that.
\end{rem}   

Let us consider some applications of soliton theory to the study
of the properties of equations of type (1).

They are based on the presentation of the  metrics (3) in the form
$$
ds^2=2z(a_3 dx^2+2a_2dxdy+a_1dy^2)-2\tau(a_4 dx^2+2 a_3 dx dy +
a_2dy^2) + 2dx dz +2dy d \tau,
$$
or
$$
ds^2=2z ds_{1}^2-2 \tau ds_{2}^2 + 2dx dz +2dy d \tau.
$$

Let us consider some  examples.

For the equation
$$
y''+H^2(x,y){y'}^3 + 3y'=0
$$
we have the metrics
$$
ds^2=2z(dx^2+H^2 dy^2)-4 \tau dx dy +2dx dz +2dy d \tau.
$$
containing two-dimensional part
$$
{ds_1}^2 = dx^2 +H^2 dy^2, \eqno (4)
$$
which is connected with theory of the KdV equation
$$
K_y +K K_x +K_{xxx}=0
$$
on the crvature $K(x,y)$ of the metric (4).

For the equation
$$
y''+{y'}^3 +3 \cos H(x,y){y'}^2 +y'=0
$$
corresponding metric is connected with integrable equation
$$
H_{xy} = sin H.
$$

     With the equations of type
$$
y''+a_4(x,y)=0
$$
a 4-dim Riemann space with metrics
$$
ds^2=-2\tau a_4 dx^2 +2dxdz+2dy d \tau
$$
and geodesics in form
$$
\ddot x=0, \quad \ddot y+a_4(x,y)(\dot x)^2=0,\quad 
\ddot \tau+a_{4y}(\dot x)^2 \tau=0
$$
$$
\ddot z-\tau a_{4x}(\dot x)^2-2\tau a_{4y}\dot x \dot y-
2a_4 \dot x \dot \tau=0
$$
are connected.

    For the  equations
$$
y''+3a_3(x,y)y'+a_4(x,y)=0
$$
    corresponding Riemann space has the metric
$$
ds^2=2(za_3-\tau a_4)dx^2-4 \tau a_3 dxdy+2dxdz+2dyd \tau
$$
and geodesics
$$
\ddot x -a_3\dot x^2=0,\quad \ddot y+2a_3 \dot x \dot y+a_4\dot x^2=0,
$$
$$
\ddot \tau -2a_3 \dot x \dot \tau-a_{3y}\dot x^2 z+(a_{4y}-2{a_3}^2-
2a_{3x})\dot x^2 \tau =0,   
$$
$$
\ddot z +2a_3 \dot x \dot z-2(a_3 \dot y +a_4 \dot x) \dot \tau+
[(a_{3x}+2{a_3}^2)\dot x^2+2a_{3y}\dot x \dot y] z-[a_{4x}\dot x^2+2(a_{4y}-
2a_3^2)\dot x \dot y+2a_{3y} \dot y^2] \tau=0.
$$

     Let us consider  the possibility to investigate the properties of 
the equations (1) using the facts from theory of embedding of Riemann 
spaces into the flat spaces.

      For the Riemann spaces of the classe one (which can be embedded into 
the 5-dimensional Eucledian space) the following conditions are fulfield
$$
R_{ijkl}=b_{ik}b_{jl}-b_{il}b_{jk}
$$    
and
$$
b_{ij;k}-b_{ik;j}=0
$$
where $R_{ijkl}$ are the components of curvuture tensor of the space with
metrics $ds^2=g_{ij}dx^idx^j$. 

    Applications of these relations for the spaces with the metrics (2)
 lead to  the conditions on the values $a_i(x,y)$.  

    For the spaces of the classe two (which admits the embedding into
6-dim Eucledean space with some signature)
the conditions are more complicated. They are
$$
R_{abcd}=e_1(\omega_{ac}\omega_{bd}-\omega_{ad}\omega_{bc})+
e_2(\lambda_{ac}\lambda_{bd}-\lambda_{ad}\lambda_{bc}),
$$
$$
\omega_{ab;c}-\omega_{ac;b}=e_2(t_c \lambda_{ab}-t_b \lambda_{ac}),
$$
$$
\lambda_{ab;c}-\lambda_{ac;b}=-e_1(t_c \omega{ab}-t_b \omega_{ac}),
$$
$$
t_{a;b}-t_{a;c}=\omega_{ac}\lambda^c_b-\lambda_{ac} \omega^c_b.
$$
and lead to the relations       
$$
\epsilon^{abcd}\epsilon^{nmrs}\epsilon^{pqik}R_{abnm}R_{cdpq}R_{rsik}=0
$$
and
$$
\epsilon^{cdmn} R_{abcd}R^{ab}_{mn}=-8e_1e_2\epsilon^{cdmn}t_{c;d}t_{m;n}
$$
   
\section{On relation with theory of the surfaces}
     
    The existence of metrics for the equations (1) may be use
for construction of the surfaces. 

    One possibility is connected with two-dimensional surfaces embedded
in a given 4-dimensional space and which are the generalization of the 
surfaces of translation. The equations for coordinates $Z^{i}(x,y)$ of 
such type ofthe surfaces are

$$
 \frac{\partial^2 Z^{i}}{\partial x \partial y} + 
\Gamma^{i}_{jk} \frac{\partial Z^{j}}{\partial x}
 \frac{\partial Z^{k}}{\partial y} =0.
$$ 

     From the condition of compatibility of this system 
it is possible to get the coefficients $a_i(x,y)$ and correspondent
 second order ODE's. 

      Another possibility for studying of two-dimensional surfaces 
in space with metrics (3) is connected with the choice of section
$$
x=x, \quad y=y, \quad z=z(x,y),\quad \tau=\tau(x,y)
$$
in space with metrics (3).

    Using the expressons
$$
dz=z_x dx+z_y dy, \quad d\tau=\tau_x dx+ \tau_y dy
$$
we get the metric 
$$
ds^2=2(z_x+za_3 - \tau a_4)dx^2 +2(\tau_x + z_y +2za_2 - 2 \tau a_3)dxdy+
2(\tau_y + za_1- \tau a_2)dy^2.
$$

    We can use this presentation for investigation of particular cases of
equations (1).

 1. The choice of the functions $z,\quad \tau$ in form
$$
z_x+za_3 - \tau a_4=0,
$$
$$
\tau_x + z_y +2za_2 - 2 \tau a_3=0
$$
$$
\tau_y + za_1- \tau a_2=0
$$
are connected with flat surfaces and are reduced at the substitution
$$
z=\Phi_x, \quad \tau=\Phi_y
$$
to the system
$$
 \Phi_{xx}=a_4\Phi_y-a_3\Phi_x,
$$
$$
\Phi_{xy}=a_3\Phi_y-a_2\Phi_x,
$$
$$
\Phi_{yy}=a_2\Phi_y-a_1\Phi_x.
$$
compatible at the conditions
$$
\alpha=0, \quad \alpha'= 0, \quad \alpha''=0.
$$

     2. The choice of the functions $z=\Phi_x, \tau=\Phi_y$
satisfying to the system of equations
$$
 \Phi_{xx}=a_4\Phi_y-a_3\Phi_x,
$$
$$
\Phi_{yy}=a_2\Phi_y-a_1\Phi_x
$$
with the coefficients $a_i(x,y)$ in form
$$
a_4=R_{xxx}, \quad a_3=-R_{xyy}, \quad a_2=R_{xyy}, \quad a_1=R_{yyy}
$$
where the function $R(x,y)$ is the solution of WDVV-equation
$$
R_{xxx}R_{yyy}-R_{xxy}R_{xyy}=1
$$
are corresponded to the equations (1) 
$$
y''-R_{yyy}y'^{3}+3R_{xyy}y'^{2}-3R_{xxy}y'+R_{xxx}=0.
$$

      The following choice of the coefficients $a_i$
$$
a_4=-2\omega,\quad a_1=2\omega,\quad a_3=\frac{\omega_x}{\omega},\quad
a_2=-\frac{\omega_y}{\omega}
$$
lead to the system
$$
 \Phi_{xx}+\frac{\omega_x}{\omega}\Phi_x+2\omega\Phi_y=0,
$$
$$
 \Phi_{yy}+2\omega\Phi_x+\frac{\omega_y}{\omega}\Phi_y=0
$$
with condition of compatibility
$$
\frac{\partial^2 \ln \omega}{\partial x \partial y}=
4\omega^2+\frac{\kappa}{\omega}
$$
wich is the  Wilczynski-Tzitzeika-equation. 

\begin{rem}  
  
    The  linear system of equations for the WDVV-equation  some surfaces 
in 3-dim projective space is determined. In the canonical form it becomes 
[13]
$$
 \Phi_{xx}-R_{xxx} \Phi_y+(\frac{R_{xxxy}}{2}-
\frac{R_{xxy}^2}{4}-\frac{R_{xxx} R_{xxy}}{2})\Phi=0,
$$
$$
 \Phi_{yy}-R_{yyy} \Phi_x+(\frac{R_{yyyx}}{2}-
\frac{R_{xyy}^2}{4}-\frac{R_{yyy} R_{xxy}}{2})\Phi=0,
$$

     The relations between invariants of Wilczynski for the linear system 
 are correspondent to the various types of surfaces. Some of them
with solutions of WDVV equation are connected.
    
\end{rem}  
\begin{rem}
From the elementary point of view the surfaces connected with the 
system of equations like the Lorenz can be constructed on such a way.
From the assumption 
$$
z=z(x,y)
$$
we get
$$
\sigma(y-x)z_x+(rx-y-zx)z_y=xy-bz.
$$
The solutions of this equation give us the examples of the surfaces
$z=z(x,y)$.
\end{rem}

\begin{rem}    

     Let us consider the system of equations
$$
\xi_{i,j}+\xi_{j,i}=2\Gamma^k_{ij} \xi_k
$$
for the Killing vectors of metrics (3).
It has the form
$$
\xi_{1x}=-a_3 \xi_1+a_4 \xi_2+(zA-t a_{4x})\xi_3+(zE+tF)\xi_4,
$$
$$
\xi_{2y}=-a_1 \xi_1+a_2 \xi_2+(zC+tD)\xi_3+(za_{1y}-tH)\xi_4,
$$
$$
\xi_{1y}+\xi_{2x}=2[-a_2 \xi_1+a_3 \xi_2+(za_{3y}-tB)\xi_3+(zG-ta_{2x})\xi_4,
$$
$$
\xi_{1z}+\xi_{3x}=2[a_3 \xi_3+a_2 \xi_4],\quad 
\xi_{1t}+\xi_{4x}=2[-a_4 \xi_3-a_3 \xi_4],
$$
$$
\xi_{2z}+\xi_{3y}=2[a_2 \xi_3+a_1 \xi_4],\quad
\xi_{2t}+\xi_{4y}=-2[a_3 \xi_3-a_2 \xi_4],
$$
$$
\xi_{3z}=0,\quad \xi_{4t}=0.
$$

    In particular case $\xi_i(x,y)$
$$
\xi_3 = \xi_4 =0, \quad \xi_i=\xi_i(x,y)
$$
we get the system of equations  
$$
\xi_{1x}=-a_3 \xi_1+a_4 \xi_2,\quad \xi_{2y}=-a_1 \xi_1+a_2 \xi_2,
$$
$$
\xi_{1y}+\xi_{2x}=2[-a_2 \xi_1+a_3 \xi_2]
$$
equivalent to the system for the z=z(x,y) and $\tau=\tau(x,y)$
\end{rem}

\begin{rem}  
      The Beltrami-Laplace operator
$$
\Delta= g^{ij}(\frac{\partial^2}{\partial x^i \partial x^j}-
\Gamma^k_{ij}\frac{\partial}{\partial x^k})
$$
can be used for investigation of the properties of the metrics (3). 
For example the equation
$$
\Delta \Psi=0
$$
has the form
$$
(ta_4-za_3)\Psi_{zz}+2(ta_3-za_2)\Psi_{zt}+(ta_2-za_1)\Psi_{tt}+\Psi_{xz}+
\Psi_{yt}=0.
$$

     Some solutions of this equation with geometry of the metrics 
(3) are connected.

     Putting the expression
$$
\Psi= \exp[zA+tB]
$$
into the equation
$$
\Delta\Psi=0
$$
we get the conditions
$$
A=\Phi_y,\quad B=-\Phi_x,
$$
and
$$
a_4 \Phi^2_y-2a_3 \Phi_x \Phi_y+
a_2 \Phi^2_x-\Phi_y \Phi_{xx}+\Phi_x \Phi_{xy}=0,
$$
$$
a_3 \Phi^2_y-2a_2\Phi_x\Phi_y+
a_1\Phi^2_x-\Phi_y \Phi_{xy}+\Phi_x\Phi_{yy}=0,
$$

    Using the equation 
$$
 g^{ij} \frac{\partial F}{\partial x^i} \frac{\partial F}{\partial x^j}=0
$$
or
$$
F_xF_z+F_yF_t-(ta_4-za_3)F_zF_z-2(ta_3-za_2)F_zF_t-(ta_2-za_1)F_tF_t=0. 
$$
it is possible to investigate the properties of isotropical surfaces
in space with metrics (3).

     In particular case the solutions of eikonal equation in form
$$
F=A(x,y)z^2+B(x,y)zt+C(x,y)t^2+D(x,y)z+E(x,y)t
$$
are existed if the following conditions are fulfilled
$$
2AA_x + BA_y - a_1 B^2 - 4a_2 AB - 4a_3 A^2=0,
$$
$$
2AB_x + BA_x + 2CA_y + BB_y - 4a_1 BC - a_2(B^2+8AC) + 4a_4 A^2=0,
$$
$$
2CB_y + BC_y + 2AC_x + BB_x - 4a_1 C^2 + a_3(B^2+8AC) + 4a_4 AB =0,
$$
$$
2CC_y + BC_x + 4a_2 C^2 + 4a_3 BC + a_4 B^2=0,
$$
$$
2AD_x + DA_x + EA_y + BD_y - 2a_1 BE - 2a_2(BD+2AE) - 4a_3 AD =0,
$$
$$
2CD_y + (BD)_x + 2AE_x + (BE)_y - 4a_1 EC - 4a_2 CD + 4a_3 AE + 4a_4 AD =0,
$$
$$
2CE_y + CE_y + DC_x + BE_x - 4a_2 CE + 2a_3(BE+2CD) + 2a_4 BD =0,
$$
$$
DD_x + ED_y -a_1 E^2 - 2a_2 DE - a_3 D^2=0,
$$
$$
EE_y + DE_x +a_2 E^2 + 2a_3 DE + a_4 D^2=0.
$$
\end{rem}
\begin{rem}
   The metric (3) has a tetradic presentation
$$
g_{ij}=\omega_{i}^a \omega_{j}^b \eta_{ab}
$$
where
$$
\eta_{ab}=\left |\begin{array}{cccc}
0 & 0 & 1 & 0 \\
0 & 0 & 0 & 1 \\
1 & 0 & 0 & 0 \\
0 & 1 & 0 & 0
\end{array} \right |.
$$ 
 
      For example we get 
$$
ds^2= 2 \omega^1 \omega^3 +2\omega^2 \omega^4
$$
where
$$
\omega^1=dx+dy,\quad \omega^2=dx+dy+\frac{1}{t(a_2-a_4)}(dz-dt),
$$
$$
\omega^4=-t(a_4dx+a_2dy),\quad \omega^3=z(a_3dx+a_1dy)+\frac{1}{(a_2-a_4)}
(a_2dz-a_4dt).
$$
and
$$
a_1+a_3=2a_2, \quad  a_2+a_4=2a_3.
$$
\end{rem}
\begin{rem}
   Some of equations on curvature tensors in space $M^4$ are connected with
ODE's. For example, the equation
$$
R_{ij;k}+R_{jk;i}+R_{ki;j}=0 \eqno{5}
$$
lead to the conditions on coefficients $a_i(x,y)$
$$
\alpha''_x+2a_3 \alpha''-2a_4 \alpha'=0,
$$
$$
\alpha_y+2a_1 \alpha'-2a_2 \alpha=0,
$$
$$
\alpha''_y+2 \alpha'_x+4a_2 \alpha''-2a_4 \alpha-2a_3 \alpha'=0,
$$
$$
\alpha_x+2 \alpha'_y-4a_3 \alpha +2a_2 \alpha'+2a_1 \alpha''=0.
$$

     The solutions of this system give us the the second 
order equations connected with the space $M^4$ with condition (5) 
on the Ricci tensor. The symplest examples are
$$
y''-\frac{3}{y}y'^2+y^3=0, \quad y''-\frac{3}{y}y'^2+y^4=0.
$$

     The conditions
$$
R_{ij;k}-R_{jk;i}=R^{n}_{ijk;n} \eqno(6)
$$
also are  interesed. They are connected with theory of nonvacuum Einstein 
spaces.
\end{rem}

\begin{rem}  

    The contruction of the Riemannian extension of
two-dimensional spaces  connected with ODE's of type (1)
can be generalized for three-dimensional spaces connected with the equations
of the form
$$
y''+c_0 +c_1x'+c_2y'+c_3{x'}^2+c_4x'y'+c_5{y'}^2+
y'(b_0+b_1x'+b_2y'+b_3{x'}^2+b_4x'y'+
b_5{y'}^2)=0,
$$
$$
x''+a_0 +a_1x'+a_2y'+a_3{x'}^2+a_4x'y'+a_5{y'}^2+
x'(b_0+b_1x'+b_2y'+b_3{x'}^2+b_4x'y'+
b_5{y'}^2)=0,
$$
where $a_i, b_i, c_i$ are the functions of variables $x,y,z$.
\end{rem}

\section
{On the Einstein-Weyl geometry}
  
    The relation between the equations in form (1) and
$$
b''=g(a,b,b') \eqno(7)
$$
with function $g(a,b,b')$ satisfying  the partial differential equation
$$
g_{aacc}+2cg_{abcc}+2gg_{accc}+c^2g_{bbcc}+2cgg_{bccc}+
$$
$$
+g^2g_{cccc}+(g_a+cg_b)g_{ccc}-4g_{abc}-4cg_{bbc} -cg_{c}g_{bcc}-
$$
$$
-3gg_{bcc}-g_cg_{acc}+ 4g_cg_{bc}-3g_bg_{cc}+6g_{bb} =0 \eqno(8)
$$
from geometrical point of view was studied by E.Cartan [14].

   He showed that Einstein-Weyl 3-folds  parameterize the families of
curves of equation (5) which is dual to the equation (1).

    Some examples of solutions of equation (5) were obtained first in [2].

    Here we consider some examples of the Einsten-Weyl spaces.

Main facts of the theory of Einstein-Weyl spaces are the following [15].

A Weyl space is smooth  manifold equipped with
a conformal metric $g_{ij}(x),$ and a symmetric connection
$$
G^k_{ij}=\Gamma^k_{ij}-\frac{1}{2}(\omega_i \delta^k_j+
\omega_j \delta^k_i -\omega_l g^{kl}g_{ij}) \eqno(9)
$$
with condition on covariant derivation
$$
D_i g_{kj}=\omega_i g_{kj}
$$
where $\omega_i(x)$ are components of vector field.

The Weyl connection $G^k_{ij}$ has a curvature tensor $W^i_{jkl}$ and the
Ricci tensor $W^i_{jil}$, is not symmetrical $W^i_{jil}\neq W^i_{lij}$
in general case.

A Weyl space satisfying the Einstein condition
$$
\frac{1}{2}(W_{jl}+W_{lj})=\lambda(x) g_{jl}(x),
$$
with some function $\lambda(x)$, is called  Einstein-Weyl space.

Let us consider some examples.

  The components of Weyl connection of 3-dim space:
$$
ds^2=dx^2+dy^2+dz^2
$$
are
$$
2G_1=\left |\begin{array}{ccc}
- \omega_{1} & - \omega_{2} & - \omega_{3} \\
\omega_{2} & - \omega_{1} & 0 \\
\omega_{3} & 0 & - \omega_{1}
\end{array} \right | ,
2G_2=\left | \begin{array}{ccc}
-\omega_{2} &  \omega_{1} & 0 \\
-\omega_{1} & -\omega_{2} & -\omega_{3}\\
0 & \omega_{3} & -\omega_{2}
\end{array} \right |,
2G_3=\left | \begin{array}{ccc}
- \omega_{3} & 0 & \omega_{1} \\
0 & - \omega_{3} & \omega_{2} \\
- \omega_{1} & - \omega_{2} & - \omega_{3}
\end{array} \right |.
$$

From the equations of Einstein-Weyl spaces
$$
W_{[ij]}=\frac{W_{ij}+W_{ji}}{2}= \lambda g_{ij}
$$
we get the system of equations
$$
\omega_{3x}+ \omega_{1z}+ \omega_1 \omega_3=0,\quad
\omega_{3y}+ \omega_{2z}+ \omega_2 \omega_3=0, \quad
\omega_{2x}+ \omega_{1y}+ \omega_1 \omega_2=0,
$$
$$
2 \omega_{1x}+ \omega_{2y}+ \omega_{3z}- \frac{ \omega_{2}^2+
\omega_{3}^2}{2}=2 \lambda,\quad
2 \omega_{2y}+ \omega_{1x}+ \omega_{3z}- \frac{ \omega_{1}^2+
\omega_{3}^2}{2}=2 \lambda,
$$
$$
2 \omega_{3z}+ \omega_{2y}+ \omega_{1x}- \frac{ \omega_{1}^2+
\omega_{2}^2}{2}=2 \lambda.
$$

Remark that the first three equations  lead to the Chazy equation [16]
$$
{R}''' + 2 R {R}''-3{R'}^2=0
$$
for the function 
$$
R=R(x+y+z) = \omega_1 + \omega_2 + \omega_3
$$
where $\omega_i= \omega_i (x+y+z)$ and in general case they 
are generalization of classical Chazy equation.

     Einstein-Weyl geometry of the metric
$g_{ij}=diag(1,-e^U,-e^U)$ and vector $\omega_i=(2U_z,0,0)$
is determined by the solutions of equation [17]
$$
U_{xx}+U_{yy}=(e^ U)_{zz}.
$$
This equation is equivalent to the equation
$$U_{\tau}=(e^{U/2})_{z}
$$
(after substitution  $U=U(x+y=\tau,z)$) having  many-valued
solutions.

    The consideration of the E-W structure for the metrics
$$
ds^2=dy^2-4dxdz-4udt^2
$$
lead to the dispersionless KP equation [18]
$$
(U_t-UU_x)_x=U_{yy}.
$$

    2. An Einstein-Weyl geometry of the four-dimensional  Minkovskii
space

$$
ds^2=dx^2+dy^2+dz^2-dt^2
$$

     The components of Weyl connection are
$$
2G_1=\left |\begin{array}{cccc}
- \omega_{1} & - \omega_{2} & - \omega_{3} & - \omega_4 \\
\omega_{2} & - \omega_{1} & 0 & 0 \\
\omega_{3} & 0 & - \omega_{1} & 0 \\
- \omega_{4} & 0 & 0 & - \omega_{1}
\end{array} \right |,\quad
2G_2=\left |\begin{array}{cccc}
-\omega_{2} &  \omega_{1} & 0 & 0 \\
-\omega_{1} & - \omega_{2} & - \omega_{3} & - \omega_4 \\
0 & \omega_{3} & -\omega_{2} & 0 \\
0 & - \omega_{4} & 0 & - \omega_{2}
\end{array} \right|,
$$
$$
2G_3=\left |\begin{array}{cccc}
- \omega_{3} & 0 & \omega_{1} & 0 \\
0 & - \omega_{3} & \omega_{2} & 0 \\
- \omega_{1} & - \omega_{2} & - \omega_{3} & - \omega_4 \\
0 & 0 & - \omega_{4} & - \omega_{3}
\end{array} \right |,\quad
2G_4=\left |\begin{array}{cccc}
- \omega_{4} & 0 & 0 & \omega_{1} \\
0 & - \omega_{4} & 0 & - \omega_{2} \\
0 & 0 - \omega_{4} & - \omega_{3} \\
- \omega_{1} & - \omega_{2} & - \omega_{3} & - \omega_4
\end{array} \right |.
$$

     The Einstein-Weyl condition
$$
W_{[ij]}=\frac{W_{ij}+W_{ji}}{2}= \lambda g_{ij}
$$
where
$$
W_{ij}=W^l_{ilj}
$$
and
$$
W^k_{ilj}=\frac{\partial G^k_{ij}}{\partial x^l}-
\frac{\partial G^k_{il}}{\partial x^j}+G^k_{in}G^n_{lj}-G^k_{jn}G^n_{il}
$$
 lead to the system of equations
$$
\omega_{3x}+ \omega_{1z}+ \omega_1 \omega_3=0,\quad
\omega_{3y}+ \omega_{2z}+ \omega_2 \omega_3=0,
$$
$$
\omega_{2x}+ \omega_{1y}+ \omega_1 \omega_2=0,\quad
\omega_{4x}+ \omega_{1t}+ \omega_1 \omega_4=0,
$$
$$
\omega_{4y}+ \omega_{2t}+ \omega_2 \omega_4=0,\quad
\omega_{4z}+ \omega_{3t}+ \omega_3 \omega_4=0,
$$
$$
3 \omega_{1x}+ \omega_{2y}+ \omega_{3z}- \omega_{4t}+ \omega_{4}^2-
\omega_{2}^2- \omega_{3}^2=2 \lambda,
$$
$$
3 \omega_{2y}+ \omega_{1x}+ \omega_{3z}- \omega_{4t}+ \omega_{4}^2-
\omega_{1}^2- \omega_{3}^2=2 \lambda,
$$
$$
3 \omega_{3z}+ \omega_{2y}+ \omega_{1x}- \omega_{4t} +\omega_{4}^3-
 \omega_{1}^2- \omega_{2}^2=2 \lambda.
$$
$$
3 \omega_{4t}- \omega_{2y}- \omega_{1x}- \omega_{3z} +\omega_{3}^2+
 \omega_{1}^2+ \omega_{2}^2=2 \lambda.
$$

\section
{On solutions of dual equations}

    Equation (8) can be written in compact form
$$
\frac{d^2 g_{cc}}{da^2}-g_{c} \frac{dg_{cc}}{da}-4 \frac{dg_{bc}}{da}+
4g_c g_{bc}-3g_b g_{cc}+6g_{bb}=0 \eqno(10)
$$
with help of the operator
$$
\frac{d}{da}=\frac{\partial}{\partial a}+
c \frac{\partial}{\partial b}+ g \frac{\partial}{\partial c}.
$$

   It has many types of 
reductions and the symplest of them are
$$
 g=c^{\alpha}\omega[ac^{\alpha-1}],\quad g=c^{\alpha}\omega[bc^{\alpha-2}],
\quad g=c^{\alpha}\omega[ac^{\alpha-1},bc^{\alpha-2}],
\quad g=a^{-\alpha}\omega[ca^{\alpha-1}],
$$
$$
\quad g=b^{1-2\alpha}\omega[cb^{\alpha-1}],
\quad g=a^{-1}\omega(c-b/a),
\quad g=a^{-3}\omega[b/a,b-ac],\quad
 g=a^{\beta/\alpha-2}\omega[b^{\alpha}/a^{\beta},
c^{\alpha}/a^{\beta-\alpha}].
$$

   To integrate a corresponding  equations let us consider some particular cases

    1. $g=g(a,c)$

From the condition (10) we get
$$
\frac{d^2 g_{cc}}{da^2}-g_{c} \frac{dg_{cc}}{da}=0 \eqno (11)
$$
where
$$
\frac{d}{da}=\frac{\partial}{\partial a}+ g \frac{\partial}{\partial c}.
$$

    Putting into (11) the relation
$$
g_{ac}=-gg_{cc}+\chi(g_c)
$$
we get the equation for $\chi(\xi),\quad \xi=g_c$
$$
\chi(\chi''-1)+(\chi'-\xi)^2=0.
$$
It has solutions
$$
\chi=\frac{1}{2}\xi^2, \quad \chi=\frac{1}{3}\xi^2
$$
 
      So we get two reductions of the equation (10)
$$
g_{ac}+gg_{cc}-\frac{g_{c}^2}{2}=0
$$
and
$$
g_{ac}+gg_{cc}-\frac{2g_{c}^2}{3}=0.
$$
\begin{rem}
   The first reduction of equation (8) is followed from its 
presentation in form 
$$
g_{ac}+gg_{cc}-\frac{1}{2}{g_c}^2+cg_{bc}-2g_b=h,
$$
$$
h_{ac}+gh_{cc}-g_c h_c+ch_{bc}-3h_b=0
$$
and was considered before.

   In particular case $h=0$ we get one equation 
$$
g_{ac}+gg_{cc}-\frac{1}{2}g_{c}^2+cg_{bc}-2g_{b}=0
$$
which is the equation (10) for the  function $g=g(a,c)$.
It can be integrate with help of Legendre transformation (see [3]).
\end{rem}

   The solutions of the equations of type 
$$
u_{xy}=uu_{xx}+\epsilon u_{x}^2
$$
was constructed in [19]. In  work of [20] was showed that they 
can be present in form
$$
u=B'(y)+\int[A(z)-\epsilon y]^{(1-\epsilon)/ \epsilon} dz,
$$
$$
x=-B(y)+\int[A(z)-\epsilon y]^{1/ \epsilon} dz.
$$

    To integrate above equations we apply the parametric representation
$$
g=A(a)+U(a,\tau), \quad c=B(a)+V(a,\tau).\eqno(11)
$$
Using the formulas
$$
g_c=\frac{g_{\tau}}{c_{\tau}}, \quad g_{a}=g_{a}+g_{\tau}\tau_{a}
$$
we get after the substitution in (10) the conditions
$$
A(a)=\frac{d B}{d a}
$$
and 
$$
U_{a \tau}-\left(\frac{V_{a} U_{\tau}}{V_{\tau}}\right)_{\tau}+
U \left(\frac{ U_{\tau}}{V_{\tau}} \right)_{\tau} -
\frac{1}{2} \frac{U_{\tau}^2}{V_{\tau}}=0.
$$

     So we get one equation for two functions $U(a,\tau)$ and $V(a,\tau)$.
Any solution of this equation are determined  the solution  of 
equation (10) in form (11).
    
     Let us consider the examples.
$$
A=B=0, \quad U=2\tau-\frac{a\tau^2}{2}, \quad V=a\tau-2\ln(\tau)
$$

   Using the representation
$$
U=\tau \omega_{\tau}-\omega,\quad V=\omega_{\tau}
$$
it is possible to obtain others solutions of this equation.

\section
{ Third-order ODE's and the Weyl-geometry}

    In the work of E.Cartan was studied the geometry of the equation
$$
b'''=g(a,b,b',b'')
$$
with General Integral in form
$$
F(a,b,X,Y,Z)=0.
$$
It was showed that there are a lot types of geometrical structures 
connected with this type of equations.

     More recently [21,22] the geometry of Third-order ODE's was considered
in context of the null-surface formalism and was discovered that in the case
 the function $g(a,b,b',b'')$ is satisfied the conditions:
$$
\frac{d^2 g_r}{da^2}-2g_{r} \frac{dg_r}{da}-3 \frac{dg_c}{da}+
\frac{4}{9}g_{r}^3 +2g_c g_r+6g_b=0 \eqno(12) 
$$
$$
\frac{d^2 g_{rr}}{da^2}-\frac{d g_{cr}}{da}+g_{br}=0 \eqno(13)
$$
where 
$$
\frac{d}{da}=\frac{\partial}{\partial a}+
c \frac{\partial}{\partial b}+ r \frac{\partial}{\partial c}+
g \frac{\partial}{\partial r}.
$$
the Einstein-Weyl geometry in space of initial dates has been
realized.
  
    We present here some  solutions of the equations (12, 13) which
is connected with theory of the second order ODE's.

   In the notations of E.Cartan we study the Third-order differential
equations  
$$
y'''=F(x,y,y',y'')
$$
where the function $F$
is satisfied to the system of conditions
$$
\frac{d^2 F_2}{dx^2}-2F_{2} \frac{dF_2}{dx}-3 \frac{dF_1}{dx}+
\frac{4}{9}F_{2}^3 +2F_1 F_2+6F_0=0 \eqno(14) 
$$
$$
\frac{d^2 F_{22}}{dx^2}-\frac{d F_{12}}{dx}+F_{02}=0 
$$
where 
$$
\frac{d}{dx}=\frac{\partial}{\partial x}+
y' \frac{\partial}{\partial y}+ y'' \frac{\partial}{\partial y'}+
F \frac{\partial}{\partial y''}.
$$

   We consider the case of equations
$$
y'''=F(x,y',y'')
$$

In this case $F_0=0$ and from the second equation we have
$$
H_{x2}+y''H_{12}+FH_{22}=0
$$
where
$$
F_{x2}+FF_{22}-\frac{F_{2}^2}{2}+y''F_{12}-2F_1=H
$$

 With the help of this the first equation give us the condition  
$$
H_x+y''(H_1-F_{11})-FF_{12}-\frac{1}{18}F_2^3-F_{x1}=0
$$

In the case
$$
H=H(F_2), \quad and \quad F=F(x,y'')
$$
we get the condition on the function $F$
$$
F_{x2}+FF_{22}-\frac{F_2^2}{3}=0
$$

    The corresponding third-order equation is
$$
y'''=F(x,y'')
$$
and it is connected with the second-order  equation 
$$
z''=g(x,z').
$$

       Another example is the solution of the system for the function 
$F=F(x, y',y'')$ obeying to the equation
$$
F_{x2}+FF_{22}-\frac{F_{2}^2}{2}+y''F_{12}-2F_1=0
$$
 
    In this case $H=0$ and we get the system of equations
$$
F_{x2}+FF_{22}-\frac{F_{2}^2}{2}+y''F_{12}-2F_1=0
$$
$$
y''F_{11}+FF_{12}+\frac{1}{18}F_2^3+F_{x1}=0
$$
with the condition of compatibility
$$
(\frac{F_2^2}{6}-F_1)F_{22}+2F_2F_{12}+3F_{11}=0
$$
\section
{Acknowledgement}

The author  thanks  the Cariplo Foundation (Centre Landau-Volta,
Como, Italy), INTAS-99-01782 Programm, the Physical Department of Roma 1 
and Lecce University for financial support and kind hospitality.

{}  


\begin{thebibliography}{}  

\bibitem{1} V. Dryuma, {\it  Application of the E. Cartan method for
studyuing of nonlinear dynamical systems}, In ''Matematicheskie 
issledovaniya, Kishinev, Stiintsa, 1987, v.92, 49-68,.
\bibitem{2} V. Dryuma, {\it Projective duality in theory of the second
order differential equations}, 
Mathematical Researches, Kishinev, Stiintsa, 1990, v.112, 93--103.
\bibitem{3} Dryuma V.S., {\it On Initial values problem in theory of the 
second order ODE's}, Proceedings of the  Workshop on Nonlineariyu,
Integrability and all that: Twenty years after NEEDS'79, 
 Gallipoli(Lecce), Italy,  July 1-July 10, 1999, ed.  M.Boiti, L.
Martina, F. Pempinelli, B.Prinari and G.Soliani, World Scientific, 
Singapore, 2000, 109-116.   
\bibitem{4} V. Dryuma, {\it On Geometry of the second order differential
equations}, Proceedings of Conference Nonlinear Phenomena, ed. K.V. Frolov,
 Moskow, Nauka, 1991, 41-48.
\bibitem{5} V. Dryuma, {\it Geometrical properties of multidimensional 
differential equations and the Finsler metrics of dynamical systems}, 
Theoretical and Mathematical Physics, Moskow, Nauka, 1994, v.99, no.2, 
241-249.
\bibitem{6} Dryuma V.S., {\it Geometrical properties of nonlinear dynamical 
systems}, Proceedings of the First Workshop on Nonlinear Physics, Le Sirenuse,
 Gallipoli(Lecce), Italy June 29-July 7, 1995, ed. E. Alfinito, M.Boiti, L.
Martina and  F. Pempinelli, World Scientific, Singapore, 1996, 83-93.   
\bibitem{7} R. Liouville, {\it Sur les invariants de  
certaines \'{e}quations diff\'{e}rentielles et sur leurs applications},  
 J. de L'\'{E}cole Polytechnique {\bf 59}, 7-76 (1889).   
\bibitem{8} A. Tresse, {\it D\'{e}termination des Invariants  
ponctuels de l'\'{E}quation differentielle ordinaire de second  
ordre: $y''=w(x,y,y') $}. Preisschriften der f\"urstlichen   
Jablonowski'schen Gesellschaft XXXII, Leipzig, S. Hirzel, 1896. 
\bibitem{9} A. Tresse, {\it Sur les invariants  
diff\'{e}rentiels des groupes continus de transformations},  
 Acta Math. {\bf 18}, 1-88 (1894).  
\bibitem{10} E. Cartan, {\it Sur les vari\'{e}t\'{e}s a  
connexion projective}, Bulletin de la Soci\'{e}t\'{e} Math\'{e}mat.   
de France {\bf 52}, 205-241 (1924).  
\bibitem{11} G.Thomsen, {\it \"{U}ber die topologischen  
Invarianten der Differentialgleichung   
$ y''=f(x,y)y'^3+g(x,y)y'^2+h(x,y) y' +k(x,y) $},  
Abhandlungen aus dem mathematischen Seminar der Hamburgischen  
Universit\"at, {\bf 7}, 301-328, (1930).  
\bibitem{12} Paterson E.M., Walker A.G.,{\it Riemann extensions}
Quart. J. Math. Oxford {\bf 3}, 19-28 (1952).
\bibitem{13} Wilczynski E.{\it Projective Differential Geometry of
Curved Surfaces}, 
Transaction of American Mathematical Society, {\bf 9}, 103-128, (1908).
\bibitem {14} Cartan E.{\it Sur une classe d'espaces de Weyl}
 Ann. Ec. Norm. Sup. {\bf 14}, 1--16, (1943).
\bibitem {15} Pedersen H., Tod K.P.
{Three dimensional Einsten-Weyl Geometry}
 Advances in Mathematics {\bf 97}, 71--109, {1993}.
\bibitem{16} Chazy J. {\it Sur les equations differentielle don'tintegrale
possede un coupure essentielle mobile}, C.R. Acad. sc. Paris, {bf 150}, 
456-458, (1910).  
\bibitem{17} Ward R.{\it Einstein-Weyl spaces and Toda fields}, 
Classical and Quantum Gravitation, {\bf 7}, 
L45-L48, (1980).  
\bibitem{18} Dunaisjski M., Mason L, Tod K.P.{\it Einstein-Weyl geometry,
the dKP equation and twistor theory}, arXiv:math. DG/0004031, 6 Apr.2000.  
\bibitem{19} Calogero F.{\it A solvable nonlinear wave equation},
Studies in Applied mathematics, {\bf LXX}, N3, 189--199, (1984).
\bibitem{20} Pavlov M.{\it The Calogero equation and Liouville type
equations}, arXiv:nlin. SI/0101034, 19 Jan. 2001.
\bibitem{21} Tanimoto M.{\it On the null surface formalism}, 
arXiv:gr-qc/9703003, 1997.
\bibitem{22}D.M.Forni, M.Iriondo, C.N.Kozameh {\it Null surface formalism
in 3D}, arXiv:gr-qc/0005120, 26 May 2000.
\bibitem{23} Tod K.P.{\it Einstein-Weyl spaces and Third-order Differential
 equations}, J.Math.Phys, N9, 2000.
\bibitem{24} S.Frittelli, C.N.Kozameh, E.T.Newman {\it Differential geometry
from differential equations}, arXiv:gr-qc/0012058, 15 Dec. 2000.

\end{thebibliography}
\enddocument